\magnification\magstep 1
\vsize=23 true cm
\topinsert\vskip 1 true cm
\endinsert
{\centerline{\bf{VISIBILITY DIAGRAMS AND EXPERIMENTAL}}}
{\centerline{\bf{STRIPE STRUCTURES IN THE QUANTUM HALL EFFECT}}}
\vskip 2 true cm
{\centerline{\bf{Yvon Georgelin$^a$, Thierry Masson$^b$ and Jean-Christophe
Wallet$^a$}}}
\vskip 2 true cm
{\centerline{$^a$Groupe de Physique Th\'eorique, Institut de Physique
Nucl\'eaire}}
{\centerline{F-91406 ORSAY Cedex, France}}
\vskip 1 true cm
{\centerline{$^b$Laboratoire de Physique Th\'eorique (U.M.R. 8627), Universit\'e 
de Paris-Sud}}
{\centerline{B\^at. 211, F-91405 ORSAY Cedex, France}}
\vskip 2 true cm
{\bf{Abstract:}} We analyze various properties of the visibility diagrams
that can be used in the context of modular symmetries and confront them to some
recent experimental developments in  the Quantum Hall Effect. We show that 
a suitable physical interpretation of the visibility diagrams which permits one 
to describe successfully the observed architecture of the Quantum Hall states 
gives rise naturally to a stripe structure reproducing some of the 
experimental features that have been observed in the study of the quantum fluctuations
of the Hall conductance. Furthermore, we exhibit new properties of the
visibility diagrams stemming from the structure of subgroups of the 
full modular group.
\vskip 1 true cm
(June 1999)
\vskip 1 true cm
LPT-99/27
\vfill\eject
{\noindent{{\bf{1. INTRODUCTION}}}}\par
\vskip 0,5 true cm
The Quantum Hall Effect (QHE) is a remarquable phenomenon occuring in a
two-dimensional electron gas in a strong magnetic field at low temperature
[1]. Since the discovery of the quantized integer [2] and fractional [3] Hall
conductivity, the QHE has been an intensive area of theoretical and
experimental investigations. The pioneering theoretical contributions [4]
analyzing the basic features of the hierarchy of the Hall plateaus have
triggered numerous works aiming to provide a better understanding of the
underlying properties governing the complicated phase diagram associated
with the quantum Hall regime together with the precise nature of the various
observed transitions between plateaus and/or focusing on a characterization
of a suitable theory.\par
It has been realized for some time that modular symmetries may well be of
interest to understand more deeply salient features of the QHE. For
instance, it has been suggested [5], [6a-b] that some properties of the phase diagram
may be explained in terms of modular group transformations. At the present
time, a fully satisfactory microscopic or effective theory for the QHE, from
which the relevant modular symmetry (if any) would come out, is
still lacking. This has motivated further studies focalized on the 
derivation of general constraints on the phase diagram coming for the 
full modular group or some of its subgroups [6a-b].\par
Some time ago, we have shown that a special subgroup of the full modular
group, namely the group $\Gamma(2)$, can be used to derive a model for a
classification of integer as well as fractional Hall states [7a-b]. We have
further shown that the constraints stemming from $\Gamma(2)$ on physically
admissible $\beta$-functions [8] give rise to a global phase diagram as well
as crossover in the various observed transitions which are in good agreement
with the present experimental observations. Here, it is worth recalling that
the classification based on the $\Gamma(2)$ symmetry [7a-b], which refines the 
Jain and Haldane ones and can be somehow 
viewed as a kind of generalization of the law of the corresponding states
proposed in [9], seems to reproduce successfully the observed hierarchical 
structure of the Hall states. This construction involves two important 
building blocks (in addition to the action of the group $\Gamma(2)$ itself) 
called the visibility diagrams. Basically, these diagrams, inherited from 
theoretical studies in arithmetics and rigidely linked to the structure of 
$\Gamma(2)$, have been shown to encode a great amount of information on the 
experimentaly observed global organization of the quantum Hall states.\par
Recently, some new experiments on the mesoscopic conductance in the quantum Hall
regime in silicon MOSFETs have been performed [10]. The main result is that the
extrema of the conductance fluctuations spread on linear trajectories in the
gate voltage $V_g$-magnetic field $B$ plane. These lines are parallels to 
lines of constant filling factor $\nu=p$ (with $p=0,1,2,...$). Moreover, a clear 
stripe structure can be observed [10] on this plane. \par
The purpose of this note is to analyze more deeply various
properties of the visibility diagrams (therefore extending our previous
investigations [7b])as well as to confront them to the
present experimental situation for the QHE. We show that a suitable physical interpretation of the
visibility diagrams that permits one to describe successfully the observed
architecture of the Quantum Hall states gives rise naturally to a stripe
structure that reproduces some of the experimental features of the
stripe structure of the conductance fluctuations in the $V_g$-$B$ plane.
This is presented in section 2 where the interesting role played by the visibility diagrams in
the description of some of the physics of the Quantum Hall regime is
emphasized. Furthermore, we exhibit in section 3 new properties of these diagrams stemming from the
structure of subgroups of the full modular group. Finally, we summarize the
results and conclude.\par
\vskip 0,5 true cm
{\noindent{{\bf{2. VISIBILITY DIAGRAMS AND STRIPE STRUCTURES}}}}\par
\vskip 0,5 true cm
Let us first recall the essential features of $\Gamma(2)$ that
will be needed in the following analyzis. The group $\Gamma(2)$ is the set of
transformations $G$ acting on the upper-half complex plane ${\cal{P}}$ which
can be written as
$$G(z)={{(2s+1)z+2n}\over{2rz+(2k+1)}},\ \ k,n,r,s\in Z \eqno(1),$$
where 
$$(2s+1)(2k+1)-4rn=1\ \ {\hbox{unimodularity condition}} \eqno(2),$$
and $z\in{\cal{P}}$ (Im$z>0$). The two generators of $\Gamma(2)$ are defined
by
$$T^2(z)=z+2, \ \ \ \Sigma(z)={{z}\over{2z+1}}  \eqno(3a,b).$$
Now to construct from $\Gamma(2)$ a model for the classification of the quantum
Hall states identify first the complex coordinate $z$ with the filling factor
$\nu=p/q$. Then, as we have shown in [7a], for a given (fixed) even
denominator metallic state $\lambda={{2s+1}\over{2r}}$, the hierarchy of the
(liquid) odd denominator states surrounding this metallic state is obtained
from the images $G^\lambda_{n,k}(0)$ and $G^\lambda_{n,k}(1)$ of $0$ and $1$
by the family of transformations $G^\lambda_{n,k}\in\Gamma(2)$ where 
$\lambda={{2s+1}\over{2r}}$ holds and $n,k$ are constrained by the
unimodularity condition (2).\par
The action of $\Gamma(2)$ on the filling factor $\nu=p/q$
can be visualized with the help of graphical representations 
called visibility diagrams whose
construction is now summarized (for a detailed construction see [7b]). Consider 
a two-dimensional square lattice whose 
vertices are indexed by a couple of positive (or zero) {\it{relatively prime}} 
integers $(q,p)$. Since $\Gamma(2)$ preserves the parity of the denominator
of any rational fraction, there are actually two ways for organizing the
vertices pertaining to this lattice, depending whether the
denominator is even or odd. \par
Consider first the case where it is even and
choose therefore a given $\lambda={{2s+1}\over{2r}}$ as a starting vertex.
Then, it is not difficult to realize that $G^\lambda_{n,k}(0)$ and 
$G^\lambda_{n,k}(1)$ which in the present framework label the Hall plateaus
surrounding the even denominator (metallic) Hall state corresponding to
$\lambda$ are all located on two parallel straigh lines forming an unbounded
left-ended stripe surrounding the vertex $\lambda$. Finally, the application
of a similar process to all even denominator fractions gives rise to a
collection of non-overlapping stripes as depicted on fig.1, each stripe
corresponding therefore to a vertex with even denominator. This visibility
diagram, hereafter called even diagram, involves naturally  
the Jain hierarchy which corresponds to the stripe associated with
$\lambda=1/2$ as it can be easily realized
by computing the successive values for $G^{1/2}_{n,k}(0)$ and $G^{1/2}_{n,k}(0)$ 
using (1) and (2). It is worth recalling some experimental 
results performed in [11-13] on metal-insulator 
transitions. It appears that the corresponding phase diagrams (see e.g.
fig.2 of ref.[11] and fig.2 of ref.[12]; see also a recent result reported in
[13]) exhibit a stripe structure which
bears some similarity with the stripe structure occuring on the even
diagram when, anticipating what will be done in a while, the
$(q,p)$ plane is identified with the magnetic field-charge carrier density
plane{\footnote\ddag{\sevenrm{at least for not too small charge carrier
density}}}. This can be easily illustrated by selecting from fig.1 the
relevant stripes which corresponds here to $\lambda=(2s+1)/2$ as shown on
fig.2 to be compared to the fig.3 of ref.[13]. We close this part of the
discussion by noting that an experimental phase
diagram of the integer QHE (charge carrier density ($\sim V_g$) versus
magnetic field) has been explicitely obtained very recently in [14] suggesting that the
vertical leftmost half stripe that is involved in the even diagram might well
be associated with some insulator state. In fact, provided the above interpretation of
the $(q,p)$ plane is actually correct, this vertical stripe (indexed by
$"1/0"$) might correspond to the type I insulator indicated in [14], whose
properties are different from the type II insulator.\par
The second diagram that can be constructed, hereafter called the odd
diagram, can be readily obtained by using a well-known 
theorem in arithmetics which states that for any relatively prime 
integers $q$ and $p$ there exist (necessarily) prime integers $a$ and $b$ such that 
$$qb-pa=\pm1  \eqno(4).$$ 
Then, for any $(q,p)$ vertex of the lattice with $q$ odd, associate the set of points
$(a,b)$ satisfying this relation. It is easy to realize that these points
are located on two parallel straight lines forming a stripe surrounding the
$(q,p)$ vertex as depicted on fig.3. We then obtain stripes for any odd
denominator fraction $q/p$. Notice that the stripes can overlap, contrary to
what happens for the even diagram {\footnote*{\sevenrm{It is not dififcult to see
that the Haldane hierarchy is involved
in the odd visibility diagram.}}}.\par
We are now in position to show that this latter diagram encodes interesting
information concerning some recent experimental results related to the
quantum Hall fluctuations
of the conductance that have been reported in [10]. The main result of this
study on the mesoscopic conductance in the Quantum Hall regime in Silicon
MOSFET is that the extrema for the conductance fluctuations spread on linear
trajectories in the $V_g$-$B$ plane parallel to constant (integer) filling
factor. Namely, the following relation 
$${{C}\over{e}}\ {{\partial V_g}\over{\partial B}}=n{{e}\over{h}},\ \ n\
{\hbox{integer}}  \eqno(5)$$
in which ${{C}\over{e}}={{\partial\rho}\over{\partial V_g}}$ is assumed to hold
($\rho$ denotes the electron density and the constant
${{C}\over{e}}$$=$$8.6\ 10^{11}$cm$^{-2}$V$^{-1}$ in [10]) has been found to be 
verified within a few percent of accuracy. Moreover, a stripe structure in the
$V_g-B$ plane has been observed, as shown on fig.4 (taken from the fig.2 of
[10]).\par
In order to confront some of these experimental results
to the present $\Gamma(2)$ framework, we
have first to exhibit a possible relation between the visibility diagram, 
the gate voltage and the applied magnetic field. This proceeds as
follows. First, recall that the filling factor $\nu={{p}\over{q}}$ can 
be expressed from its very definition as
$$\nu={{N_c}\over{N_\Phi}}  \eqno(5)$$
in which $N_c$ is the number of charge carriers (the electrons in the present
case) and $N_\Phi=$$BS{{h}\over{e^2}}$ ($S$ is the device area) is the number
of unit flux. Then observe that the action of the operator $T^2$ (3.a), a Landau shift 
type operator, on any vertex $(q,p)$ of the visibility
diagram gives rise to a vertical shift ($T^2:(q,p)\to(q,p+2q)$) whereas the
action on any vertex of $\Sigma$ (3.b) which is the flux attachement
operator produces a horizontal shift ($\Sigma:(q,p)\to(q+2p,p)$). Next,
observe that the charge carrier density happens to be proportional to the
gate voltage $V_g$ in the experiments reported in [10] and that the flux
attachement is obtained by magnetic field variation. Bringing this all
together, this suggests to identify naturally (up to dimensionful factors)
the horizontal (resp. vertical) axis on the visibility diagram with the
$B$-axis (resp. $V_g$-axis).\par
Owing to the above identification, it is now possible to perform a
comparison between the stripe structure occuring in the $V_g-B$ plane 
observed in [10] (represented by the grey areas associated with the
plateaus) and the one stemming from the odd diagram.
These structures are depicted respectively on fig.4 and fig.5.
For the sake of clarity, we have only represented on fig.5 the stripes
corresponding to the integer filling factors considered in [10]. We observe a
good qualitative agreement between both structures. Each grey area on fig.4
corresponding to a given (integer) plateau appears to be bounded by two
parallel straight lines (at least in the considered range for $V_g$ and
$B$). Taking into account the method for constructing the odd diagram, it
seems natural to identify each grey area associated with a given integer filling factor
with the corresponding stripe on fig.5.\par
If this latter identification is physically correct, it is possible to obtain from the odd diagram 
further information on the Hall conductance as a function of $V_g$
(uppermost onset of fig.4) by adapting to the present situation, for which
$B$ is fixed while $V_g$ ($N_c$) varies, the argument that we used in [7b] to
obtain a resistivity plot (Hall resistivity versus $B$) fitting
well with the experimental plot. This is straighforwardly
achieved by simply assuming that the vertical width of any stripe on
fig.5, defined by the intercept of any vertical line with that stripe, is
proportional to the width of the corresponding plateau. The subsequent
analysis is then very similar to the one that we described in [7b]. We have
found that the resulting conductivity-$V_g$ plot agrees qualitatively with
the one depicted on fig.4.\par
At this point, the analysis suggests that the proposed physical interpretation of the
odd diagram is consistent with the experimental observations corresponding
to the {\it{integer}} QHE. Proving that this diagram encodes some relevant
properties of both integer and fractional QHE (giving therefore some
global information on the phase diagrams) would require further
confrontation with experiments exploring the fractional quantum Hall regime
(and/or for higher magnetic field). In this regime, it is obvious from the
very construction of the odd diagram that one (experimentaly
testable) prediction of the present scheme is the occurence of branching
tree-like structures [7b] among the stripes in the $V_g-B$ plane similar to the
one appearing on fig.3.\par 
Let us now consider the quantum Hall fluctuations. Recall that these
fluctuations are observed in the transition regions between plateaus which
narrow as the temperature decreases while the fluctuations grow and sharpen.
Obviously, a full confrontation of the experimentaly verified relation (5)
to the present $\Gamma(2)$ framework would require to have at hand a
detailled model for the quantum Hall fluctuations 
{\footnote\dag{\sevenrm{Note that relation (5) seems to
contradict the predictions stemming from the non-interacting models.}}}
providing a satisfactory explanation for the origin of the observed
behaviour. Nevertheless, keeping in mind the above analysis of the stripe
structures, it seems plausible to conjecture that, for a given transition
$\nu=n_1\to\nu=n_2,\ n_{1,2}\in N$ (extending to $n_{1,2}\in Q$ if the
present scheme applies to the fractional QHE), the directions of the two coexisting
families of straight lines involving the extrema of the fluctuations (which
are observed in [10] for the transitions $0\to1$, $1\to2$, $2\to3$, 
$3\to4$) are given by the directions defined by
the corresponding stripes involved in that transition.\par
\vskip 0,5 true cm
{\noindent{{\bf{3. MORE ON VISIBILITY DIAGRAMS AND DISCUSSION}}}}\par
\vskip 0,5 true cm
From the above analysis, it appears that the visibility diagrams may well
encode some physically relevant features of the QHE which motivates a deeper
investigation of mathematical properties underlying their structure. This is what we consider now.\par 
First, we point out that the odd diagram is in fact the superposition of two "more
elementary" visibility diagrams. To see that, consider separately odd and even
{\it{numerator}} filling factors (with odd denominator) and apply
the method for constructing the odd visibility
diagram that has been described in section 2. Doing this, one obtains
the two new diagrams represented on fig.6 and fig.7 corresponding respectively
to odd and even {\it{numerator}} filling factors (hereafter called
respectively odd/odd and even/odd diagrams). Then, it can be easily
realized that the superposition of these two latter diagrams gives rise to
the odd visibility diagram. Note that the stripes appearing in these two
diagrams do not overlap as it is the case for the even diagram. Furthermore, observe 
that this latter diagram is related to the even/odd diagram 
through a symmetry around the $p=q$ axis which corresponds to the
action of an operation belonging to the full modular group $\Gamma(1)$ but
{\it{not}} to $\Gamma(2)$.\par
Let us study more closely the action of modular transformations
pertaining to $\Gamma(1)$ on the even, even/odd
and odd/odd diagrams. Some remarks are in
order. On one hand, it can be easily seen that the 
action of any $G\in\Gamma(1)$ preserves the
arithmetic relation (4) ruling the whole construction of these diagrams.
As a consequence, the action of $\Gamma(1)$ maps the stripe structure of
each of the diagrams into another one or possibly a sub-structure. 
On the other hand, any $G\in\Gamma(2)$ maps each of these three diagrams
into itself simply because $\Gamma(2)$ preserves the even or odd character
of both numerator and denominator involved in the filling factor. In other
words, the whole structure of each diagram remains invariant under
$\Gamma(2)$. In fact, it appears that $\Gamma(2)$ is the largest subgroup of $\Gamma(1)$
leaving invariant each of the 3 diagrams. To see that, consider the action of the coset
group $\Gamma(1)/\Gamma(2)$ on these diagrams. It is known
in the mathematical litterature [15] that this coset group involves 6
elements whose corresponding representatives in $\Gamma(1)$ can be
choosen as:
$$I=\pmatrix{1&0\cr0&1\cr},\ \ U=\pmatrix{1&1\cr0&1\cr},\ \
V=\pmatrix{0&-1\cr1&0\cr}  \eqno(6a,b,c)$$
$$W=\pmatrix{1&0\cr1&1\cr},\ \ P=\pmatrix{0&-1\cr1&1\cr},\ \
P^2=\pmatrix{-1&-1\cr1&0\cr}  \eqno(6d,e,f),$$
from which it can be verified that any of the 6 coset group elements maps
a visibility diagram into a sub-diagram included in another diagram.
Finally, using (6) together with the definition of
$\Gamma(2)$, it is easy to prove that the even and odd diagrams (depicted
respectively on fig.1 and fig.2) are
invariant under the action of another subgroup of $\Gamma(1)$ generated by
$U$ given by (6b) and $\Gamma(2)$. This subgroup is nothing but
$\Gamma_0(2)$, which has been proposed [6b] (see also second of ref. [6a]) as another candidate for a discrete
symmetry group underlying the physics of the QHE.\par
Note that the constraints from $\Gamma_0(2)$ on the renormalization group
flow in a two-parameter scaling framework have been examined in [6b]. The
resulting flow diagram (phase diagram) has been shown to exhibit a specific
feature. In fact, consistency with the present experimental observations 
requires the occurence for the $0\to1$
transition {\footnote\dag{\sevenrm{Recall that, as usual, the whole flow diagram is
obtained by applying successive $\Gamma_0(2)$ transformations to the $0\to1$
"template" transition}}} of a critical point at
$\sigma_{xy}=\sigma_{xx}=1/2$ which appears as a pole of the corresponding
$\beta$-function. This stemms from the existence of a fixed point of
$\Gamma_0(2)$ in its fundamental domain at $z_0=(1+i)/2$ (recall that in this
framework one has $z=\sigma_{xy}+i\sigma_{xx}$ which parametrizes the
conductivity plane). The situation is different
in the $\Gamma(2)$ case, as shown recently in [8]: There is no critical point
showing up as a pole (at finite distance in the conductivity plane) 
of the corresponding $\beta$-function but consistency with the two-parameter
scaling hypothesis seems to require the
occurence in each allowed transition of a temperature-independant 
point that might be identified with the crossing point appearing 
in the crossover of the observed transitions [8].\par
Let us summarize the results involved in this paper. We have analyzed
various properties of the visibility diagrams which are related to the
modular symmetries. In particular, we have shown that the observed structures
occuring in the (gate voltage-magnetic field) data, at least in the integer Quantum
Hall regime, may well be encoded in one visibility diagram, namely the odd
one. We have also indicated an experimental way to test its possible relevance
to the fractional QHE. Furthermore, we have conjectured that for a given transition
$\nu=n_1\to\nu=n_2,\ n_{1,2}\in Q$ the directions of the two coexisting
families of straight lines involving the extrema of the fluctuations (
observed in [10] for the transitions $0\to1$, $1\to2$, $2\to3$, 
$3\to4$) are given by the directions defined by the corresponding 
stripes of the odd diagram involved in that transition.

\vfill\eject
{\noindent{\bf{REFERENCES}}}\par
\vskip 1 true cm
\item{} 1) The Quantum Hall Effect, 2nd ed., R.E. Prange and S.M. Girvin
eds. (Springer-Verlag, New-York) 1990. See also Perspectives in Quantum Hall
Effect, S.D. Sorma and A. Pinczuk eds. (Wiley, New-York) 1997.
\item{} 2) K. von Klitzing, G. Dorda and M. Pepper, Phys. Rev. Lett. 45
(1980) 494.
\item{} 3) D.C. Tsui, H.L. St\"ormer and A.C. Gossard, Phys. Rev. Lett. 48
(1982) 1559.
\item{} 4) R.B. Laughlin, Phys. Rev. Lett. 50 (1983) 1385; F.D. Haldane,
Phys. Rev. Lett. 51 (1983) 605; B.I. Halperin, Phys. Rev. Lett. 52 (1984)
1583.
\item{} 5) C.A. L\"utken and G.G. Ross, Phys. Rev. B45 (1992) 11837, Phys.
Rev. B48 (1993) 2500.
\item{} 6a) C.A. L\"utken, Nucl. Phys. B396 (1993) 670; C.P.
Burgess and C.A. L\"utken, On the implication of discrete symmetries for the
$\beta$-function of Quantum Hall System, cond-mat/9812396 and references
therein. See also E. Fradkin and S. Kivelson, Nucl. Phys. B474 (1996) 543.
\item{} 6b) B.P. Dolan, J. Phys. A: Math. Gen. 32 (1999) L243; B.P. Dolan,
Modular invariance, universality and crossover in the QHE, cond-mat/9809294.
\item{} 7a) Y. Georgelin and J.C. Wallet, Phys. Lett. A224 (1997) 303.
\item{} 7b) Y. Georgelin, T. Masson and J.C. Wallet, J. Phys. A:Math. Gen.
30 (1997) 5065.
\item{} 8) Y. Georgelin, T. Masson and J.C. Wallet, $\Gamma(2)$ modular
symmetry, renormalization group flow and the Quantum Hall Effect,
cond-mat/9906193.
\item{} 9) S. Kivelson, D.H. Lee and S.C. Zhang, Phys. Rev. B46 (1992) 2223.
\item{} 10) D.H. Cobden, C.H. Barnes and C.J.B. Ford, Quantum Hall
Fluctuations and evidence for charging in the Quantum Hall Effect,
cond-mat/9902154.
\item{} 11) S.V. Kravchenko, W. Mason and J.E. Furneaux, Phys. Rev. Lett. 75
(1995) 910.
\item{} 12) S.V. Kravchenko et al., Phys. Rev. B51 (1995) 7038.
\item{} 13) M. Hilke et al., experimental phase diagram of the integer
quantized Hall effect, cond-mat/9906212.
\item{} 14) D.N. Sheng and Z.Y. Weng, phase diagram of the integer Quantum
Hall Effect, cond-mat/9906261.
\item{} 15) See in D. Mumford, Tata Lectures on Theta functions, vol. I-III,
(Birkhauser, Boston, Basel, Stuttgart) 1983.

\end